\documentclass[aps,twocolumn,superscriptaddress,floatfix,nofootinbib]{revtex4-2}
\usepackage{orcidlink}
\usepackage{amsmath,amssymb,graphicx,bm}
\usepackage{caption}
\usepackage{hyperref}
\hypersetup{
colorlinks=true,
linkcolor=blue,
citecolor=blue,
urlcolor=blue
}
\usepackage{subcaption}
\usepackage{ragged2e}
\usepackage{etoolbox}

\makeatletter
\patchcmd{\@makecaption}
  {\raggedright}
  {\justifying}
  {}{}
\makeatother
\usepackage{mwe}
\usepackage{graphicx,wrapfig,lipsum}
\usepackage[export]{adjustbox}
\usepackage[utf8]{inputenc}
\usepackage[dvipsnames]{xcolor}
\usepackage{tabularx} 
\usepackage{amsmath}  
\usepackage{graphicx} 
\usepackage{tikz}

\begin{document}

\title{Anomaly-Induced Hybrid Bulk Electromagnetic Mode in Weyl Semimetals}

\author{Subrahmanyam D \orcidlink{0009-0004-2477-3937}}
\email{subrahmanyam20@iiserb.ac.in}
\affiliation{Department of Physics, Indian Institute of Science Education and Research, Bhopal 462066, India}
\author{Suhas Gangadharaiah \orcidlink{0000-0001-7834-9438}}
\email{suhasg@iiserb.ac.in}
\affiliation{Department of Physics, Indian Institute of Science Education and Research, Bhopal 462066, India}
\author{E. G. Mishchenko \orcidlink{0009-0004-9888-7128}}
\email{mishch@physics.utah.edu}
\affiliation{Department of Physics, University of Utah, Salt Lake City, Utah 84112, USA}
\date{\today}

\begin{abstract}
Collective modes provide direct fingerprints of quantum matter. We predict a previously unidentified hybrid bulk electromagnetic mode in Weyl semimetals arising from the interplay between the chiral anomaly and the orientation of its associated chiral magnetic response relative to the direction of the wave-vector. When the anomaly-induced chiral magnetic current has a component along the propagation direction, oscillations of valley imbalance hybridize with plasmonic charge oscillations, producing a linearly dispersing mode that undergoes avoided crossing with the bulk plasmon, producing a hybrid bulk excitation absent in ordinary metals. The hybrid mode provides a direct signature of Weyl semimetals and a probe of the chiral anomaly and its associated chiral magnetic effect, with observable features in electron energy-loss spectroscopy. Studying this interplay can uncover various optical and electronic properties of Weyl semimetals.
\end{abstract}

\maketitle

\paragraph*{Introduction.---}
Weyl semimetals (WSMs) are three-dimensional topological materials whose low-energy quasiparticles behave as massless Weyl fermions~\cite{Weyl1929,Balents2011,Wan2011, Burkov2011, Armitage2018,Murakami2007}. Their band structure hosts pairs of nondegenerate band-touching points, known as Weyl nodes, which act as monopoles of Berry curvature in momentum space. Each node carries a definite chirality, and the separation of nodes in momentum or energy space endows WSMs with a range of unconventional topological electromagnetic responses absent in ordinary metals, such as the intrinsic anomalous Hall effect \cite{Burkov2014,Shekhar2018,Bednik2016} and large Kerr and Faraday rotations without an applied magnetic field \cite{Kargarian2015,Ghosh2023}. A defining hallmark of WSMs is the \textit{chiral anomaly}---the nonconservation of chiral charge in the presence of parallel electric and magnetic fields $(\mathbf{E}\!\cdot\!\mathbf{B}\neq\!0)$~\cite{Adler1969,Bell1969,Son2013,dosReis2016,PhysRevB.99.075114}, which gives rise to an induced current along $\mathbf{B}$, known as the chiral magnetic effect (CME)~\cite{Fukushima2008,Kharzeev2014,Li2016,PhysRevB.93.201202}. This anomaly-driven response underlies a variety of macroscopic phenomena such as negative longitudinal magnetoresistance \cite{Son2013,Burkov2015}, anisotropic magnetoconductance, planar Hall effect and planar Nernst effect~\cite{Nandy2017,Zhang2021}.

 Beyond dc transport, the interplay of the chiral anomaly with collective charge dynamics profoundly modifies the electromagnetic modes of WSMs. In ordinary three-dimensional metals, the bulk electromagnetic spectrum consists of a longitudinal plasmon mode and transverse plasmon polariton modes. In contrast, an undoped Weyl semimetal cannot sustain long-wavelength plasmon oscillations because the density of states vanishes at the Weyl nodes, thereby suppressing collective charge density oscillations \cite{Lv2013}. However, in doped WSMs, these longitudinal and transverse bulk excitations become intrinsically anisotropic due to the gyrotropic response induced by the Weyl node separation vector in momentum space $\mathbf{b}$. As a result, the collective modes depend on the relative orientation between the propagation vector $\mathbf{q}$ and the node-separation vector, giving rise to distinct parallel configuration ($\mathbf{q}\!\parallel\!\mathbf{b}$) and perpendicular configuration $(\mathbf{q}\!\perp\!\mathbf{b})$~\cite{pellegrino2015helicons}. Early works analyzed conventional bulk plasmons and polaritons in topological semimetals without accounting for anomaly-induced dynamics~\cite{Lv2013,PhysRevB.99.075137}, while subsequent studies incorporated Berry curvature and axion electrodynamics to describe helicons and magnetoplasmons in static magnetic fields~\cite{pellegrino2015helicons,PhysRevB.104.205141,PhysRevLett.120.037403}. Chiral anomaly--induced corrections to the plasmon frequency were previously studied assuming a steady state chiral charge imbalance generated by $\mathbf{E}\cdot\mathbf{B}$ field~\cite{PhysRevB.91.035114}. Such treatments neglect the role of the orientation of the anomaly-activating magnetic field. Gorbar \textit{et al.} reported chiral magnetic (psuedomagnetic) plasmons, characterized by gapped modes that depend on the chiral shift parameter $\mathbf{b}$ when the Chern-Simons contribution is taken into account \cite{PhysRevLett.118.127601,Gorbar2017CMP}. These modes originate from the coupling between oscillations in electric and chiral current densities and are ultimately driven by the dynamical chiral separation effect. Song \textit{et al.} introduced an acoustic collective mode, chiral zero sound (CZS), arising purely from the CME, wherein the chiral anomaly enters implicitly via the CME current, producing a chemical potential imbalance between opposite chiral valleys \cite{PhysRevX.9.021053}. This mode occurs only in Weyl semimetals with multiple pairs of nodes and remains decoupled from charge dynamics. In contrast, we treat the chiral anomaly as a dynamical process by explicitly incorporating the $\mathbf{E}\!\cdot\!\mathbf{B}$ source term into the continuity equation, allowing the chiral imbalance to oscillate in phase with the collective electron density. Within this framework, we uncover a previously unidentified bulk electromagnetic excitation in Weyl semimetals: an anomaly-induced hybrid mode generated by the coupling between plasmonic charge oscillations and dynamical chiral imbalance oscillations. We show that the existence and character of this hybrid electromagnetic mode are controlled by the orientation of the associated chiral magnetic response relative to the wave vector, leading to a qualitatively new bulk collective excitation absent in earlier works.

In this work, we develop a semiclassical electromagnetic framework that treats the chiral charge density as a dynamical variable that oscillates in phase with the collective electron charge density in the presence of the chiral anomaly. 
\begin{figure}[t]
\centering
\includegraphics[width=\columnwidth]{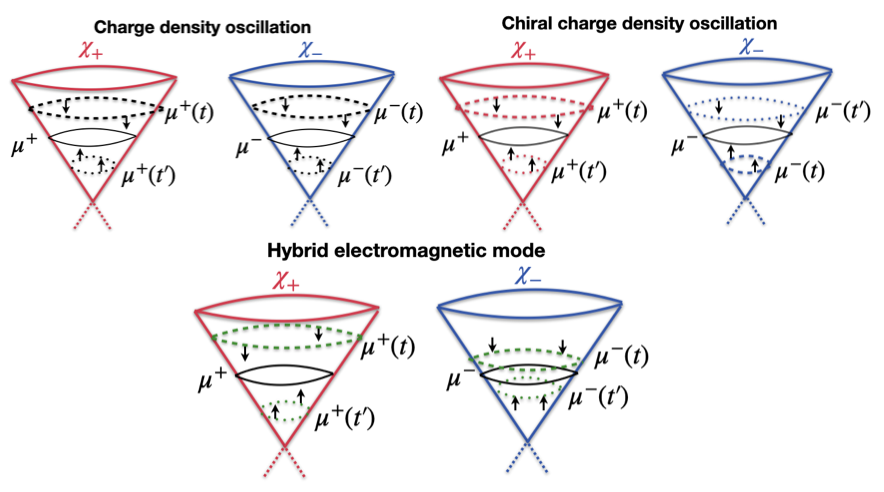}
\caption{\justifying
Schematic illustration of the anomaly-induced hybrid bulk electromagnetic mode in a Weyl semimetal in terms of breathing of the Fermi surfaces. Top left: in-phase breathing at opposite Weyl nodes, corresponding to the conventional plasmonic charge-density oscillation. Top right: antiphase breathing under $\mathbf{E}(t)\!\cdot\!\mathbf{B}_0 \neq 0$, corresponding to a chiral charge-density oscillation. Bottom: for $\mathbf{B}_0\!\cdot\!\mathbf{q}\neq 0$ and $\omega_p \simeq \omega_{\rm anom}$, coupling between the two oscillations yields a hybridized collective electromagnetic mode.}
\label{fig:schematic}
\end{figure}
Within this framework, the orientation of the anomaly-activating magnetic field can be treated explicitly. We show that when the magnetic field has a nonzero component along the direction of the wave-vector, a linearly dispersing electromagnetic mode appears in the long-wavelength limit, which smoothly hybridizes with the conventional bulk plasmon at larger wave vectors. In contrast, when the magnetic field is transverse to the direction of the wave-vector, the linear mode and the associated hybridization are absent. We further find that the damping of the anomaly-induced hybrid bulk mode depends sensitively on the orientation of the magnetic field relative to the direction of the wave-vector, in both the parallel and perpendicular configurations, and we analyze this dependence in detail. A schematic illustration of the resulting anomaly-induced hybrid electromagnetic mode is shown in Fig. \ref{fig:schematic}. This mechanism is distinct from the undamped gapless plasmons reported in tilted Dirac semimetals~\cite{agarwal2020novel}, which originate from electron--hole pocket coupling rather than chiral anomaly dynamics. This mode also differs from the linearly dispersing surface plasmon modes of Fermi-arc states~\cite{Jafaric}, which are two-dimensional boundary excitations that depend on surface boundary conditions and disappear when propagation is parallel to the arc direction. In contrast, the present mode is a three-dimensional bulk electromagnetic excitation whose velocity is explicitly controlled by the background magnetic field through the anomaly-induced velocity scale, providing a clear experimental distinction.
\paragraph*{Model and setup.---}
 We consider a doped Weyl semimetal with broken time reversal symmetry characterized by a finite Weyl-node separation vector $\mathbf{b}$ set by the internal magnetic order. Without loss of generality, we take the wave-vector $\mathbf{q}$ along the $y$ axis, corresponding to a field profile $\mathbf{E} = (E_x,E_y,E_z)e^{-i\omega t + i q y}$, and study two configurations: (i) the \textit{parallel configuration}, where $\mathbf{b}\parallel\mathbf{q}$ (node separation along $y$), and (ii) the \textit{perpendicular configuration}, where $\mathbf{b}\perp\mathbf{q}$ (node separation along $z$). To activate the chiral anomaly and the associated chiral magnetic response, we applied a classically
weak static magnetic field confined to the $xy$ plane, $\mathbf{B}_0 = B_{0x}\,\hat{x} + B_{0y}\,\hat{y}$, such that the Landau quantization is negligible. The magnetic field is oriented at an arbitrary angle $\theta$ with respect to the direction of wave-vector. This setup allows us to capture the anomaly-induced collective electromagnetic mode and the corresponding hybridization behavior in both geometries within a unified framework. The magnetic field enters only through the chiral magnetic and chiral anomaly terms, which modify the charge and current dynamics without altering the band structure. In the following section, we develop the semiclassical formalism and solve the coupled kinetic–Maxwell equations to reveal the emergence of this anomaly-induced hybrid electromagnetic mode.


 
 \paragraph*{Formalism and Key Equations.---}
Before introducing the anomaly-induced effects, we first recall the electromagnetic response of a doped Weyl semimetal in the absence of a static magnetic field. The Maxwell equations in momentum space can be written in matrix form as
\begin{equation}
    \label{eq:Mmatrix}
    \mathcal{M}^{\parallel/\perp}_{\ell m}(\omega,q)\,\mathbf{E}_{m} = 0,
\end{equation}
where $\mathcal{M}^{\parallel/\perp}_{\ell m} = c^{2}\!\left(q^{2}\delta_{\ell m}-q_{\ell}q_{m}\right) - \omega^{2}\varepsilon^{\parallel/\perp}_{\ell m}$ represents the electromagnetic response tensor for the parallel and perpendicular propagation configurations defined earlier \cite{pellegrino2015helicons}. The zeros of $\det[\mathcal{M}^{\parallel/\perp}]$ determine the conventional bulk electromagnetic modes of the Weyl semimetal in the absence of a magnetic field.

In the presence of a weak static magnetic field $\mathbf{B}_0$ confined to the $xy$ plane, the current response at a given Weyl node $\alpha$ is modified by Berry curvature and chiral magnetic contributions. The total current density at node $\alpha$ can be expressed as \cite{Sukhachov2022PRL}
\begin{align}
    \label{eq:Jalpha}
    &\mathbf{j}_\alpha(t,\mathbf{r}) = \sigma_\alpha(\omega)\,\mathbf{E}(t,\mathbf{r})
    + \mathbf{E}(t,\mathbf{r}) \times \boldsymbol{O}_\alpha-D_\alpha\mathbf{\nabla}N_\alpha(t,r)\notag\\
    &\hspace{5.3cm}- \mathbf{v}_{\Omega,\alpha}\, N_\alpha(t,\mathbf{r}).
\end{align}
Here $\sigma_{\alpha}(\omega)$ denotes the ac Drude conductivity, the second term represents the anomalous Hall–like current arising from the Berry curvature, the third and last term correspond to the diffusion current and the chiral magnetic current driven by the static magnetic field, respectively. Note that, $
\sum_{\zeta} \boldsymbol{O}_{\zeta}=\frac{2e^2}{\pi \hbar}\,\mathbf{b}$, where $\sum_{\zeta}$ is the sum over Weyl nodes $\alpha$ and $\beta.$ The diffusion constant is given by $D_\alpha=v_{F,\alpha}^2\tau_{\alpha,\alpha}/3$, where $\tau_{\alpha,\alpha}$ is the intranode scattering time and $v_{F,\alpha}$ is the Fermi velocity at node $\alpha$. For smooth disorder or weak interface roughness, the associated scattering processes involve only small momentum transfer $|\Delta \mathbf{k}| \ll |\mathbf{2b}|$. Consequently, large-momentum-transfer internode scattering is strongly suppressed. The anomalous velocity associated with the chiral magnetic response is given by $\mathbf{v}_{\Omega,\alpha} = \chi_{\alpha}e\mathbf{B}_0/(4\pi^{2}\hbar^{2}c\,\nu_{\alpha})$ where $\chi_\alpha=\pm1$ denotes the chirality. This velocity couples the perturbed electron density $N_{\alpha}$ at node $\alpha$ to the external magnetic field, giving rise to the dynamical chiral magnetic effect.
 Substituting this current into Maxwell’s equations yields 
\begin{align}
\begin{pmatrix}
\mathcal{M}^{\parallel/\perp}
\end{pmatrix}
\begin{pmatrix}
E_x \\ E_y \\ E_z
\end{pmatrix}
= \begin{pmatrix}
-\frac{4\pi i \omega}{c^2}\sum_\zeta v_{\Omega,\zeta,x} N_\zeta \\
-\frac{4\pi i \omega}{c^2}\sum_\zeta v_{\Omega,\zeta,y} N_\zeta \\
0
\end{pmatrix}\begin{pmatrix}
E_x \\ E_y \\ E_z
\end{pmatrix},
\label{eq: nonhomME}
\end{align}
where the detailed derivation is provided in the Supplementary Material \cite{Supplemental}. Eq.\eqref{eq: nonhomME} describes the complete bulk electromagnetic response in the presence of a static magnetic field. In this case, Maxwell’s equations become inhomogeneous, the right hand side of eq.\eqref{eq: nonhomME} sourced by the node-resolved perturbed carrier density imbalance \(N_\alpha-N_\beta\). 

\begin{figure}[t]

    \begin{subfigure}[b]{0.235\textwidth}
        \includegraphics[width=\linewidth]{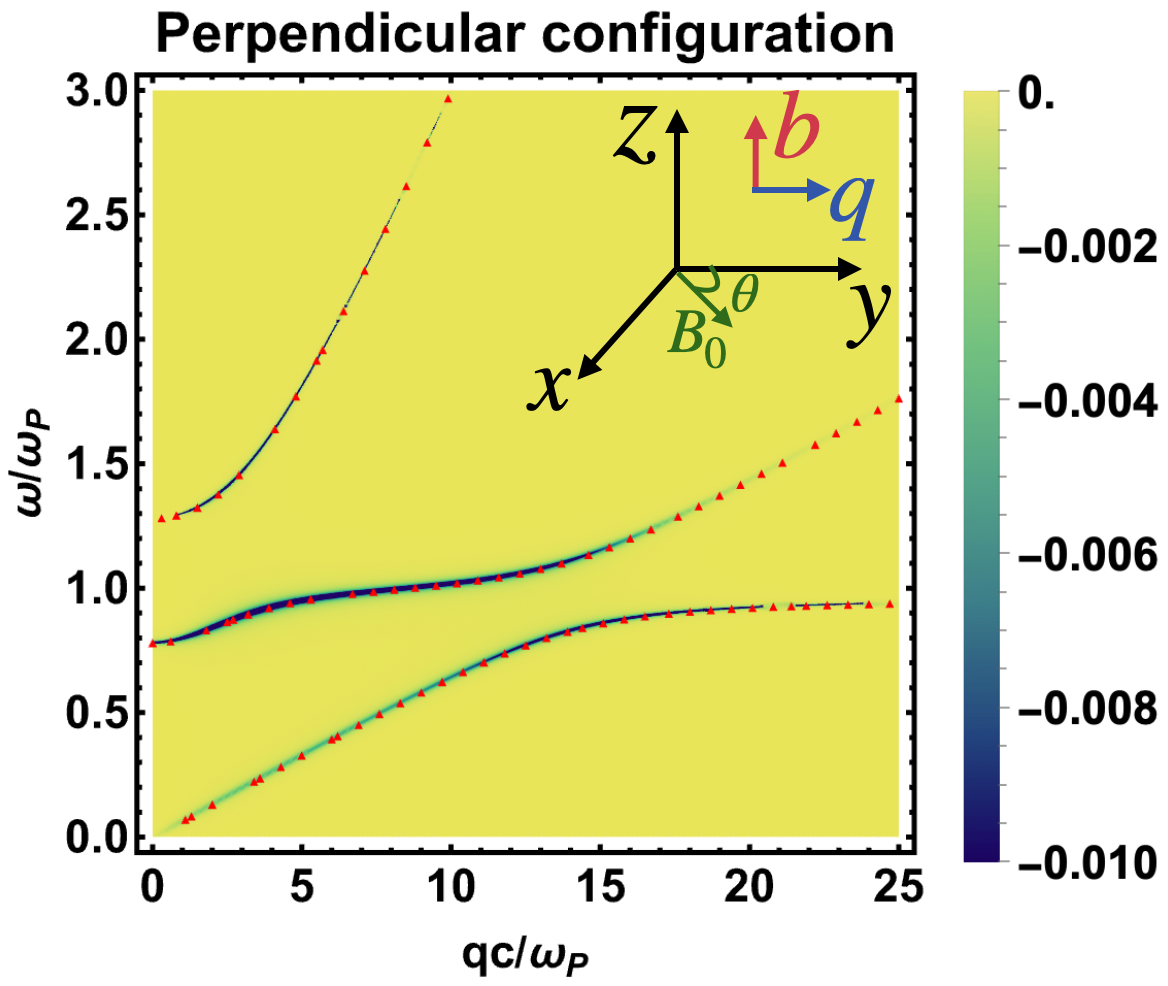}
        \caption{$\theta=30^\circ$}
        \label{fig:1a}
    \end{subfigure}
    \hfill
    \begin{subfigure}[b]{0.235\textwidth}
        \includegraphics[width=\linewidth]{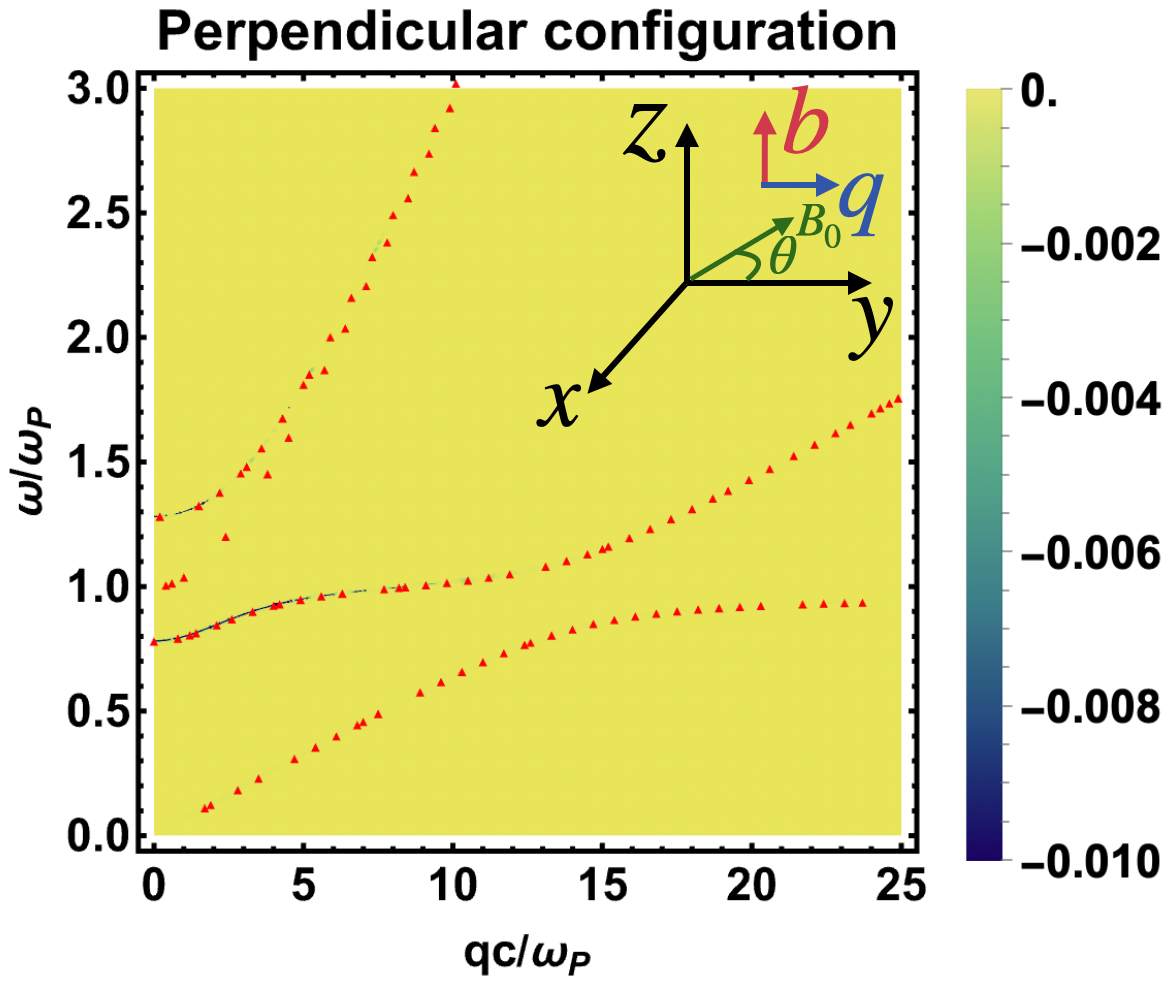}
        \caption{$\theta=30^\circ$}
        \label{fig:1b}
    \end{subfigure}

    \vspace{0.15cm}

    \begin{subfigure}[b]{0.235\textwidth}
        \includegraphics[width=\linewidth]{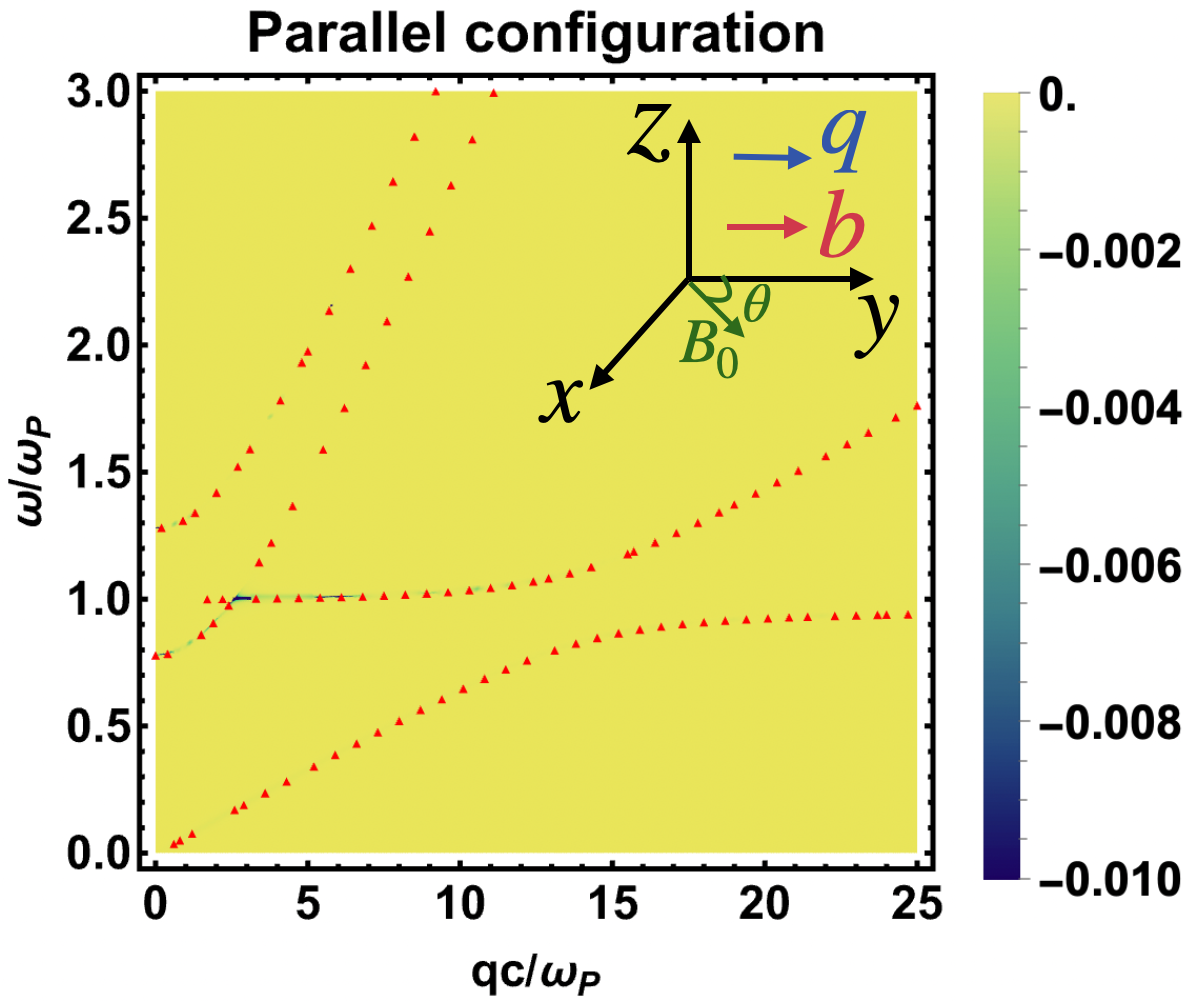}
        \caption{$\theta=30^\circ$}
        \label{fig:1c}
    \end{subfigure}
    \hfill
    \begin{subfigure}[b]{0.235\textwidth}
        \includegraphics[width=\linewidth]{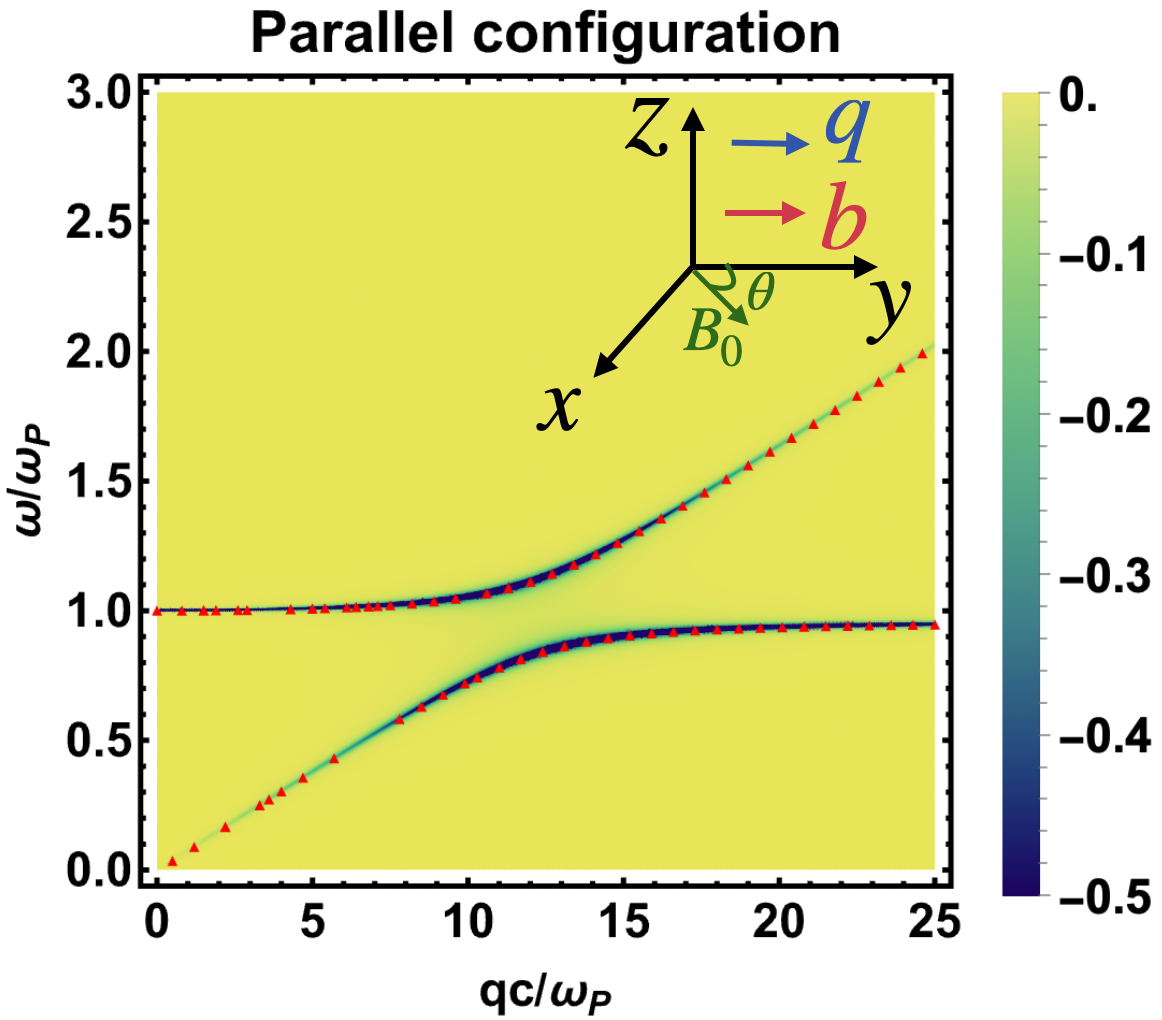}
        \caption{$\theta=0^\circ$}
        \label{fig:1d}
    \end{subfigure}

   \caption{\justifying
Bulk electromagnetic modes (red triangles) overlaid on the loss-function spectra $L(\omega,q)$ for perpendicular and parallel configurations. Peak in the loss function (blue) indicate the presence of collective modes, with their sharpness reflecting the degree of damping, while yellow regions correspond to negligible loss. Figure~\ref{fig:1a} shows $\mathbf{B_0}$ in the $x$--$y$ plane at $\theta=30^\circ$ with respect to $\hat{y}$, while Fig.~\ref{fig:1b} corresponds to $\mathbf{B_0}$ in the $y$--$z$ plane at the same angle. Figures~\ref{fig:1c} and \ref{fig:1d} (decoupled longitudinal sector) show $\mathbf{B_0}$ in the $x$--$y$ plane at $\theta=30^\circ$ and $0^\circ$, respectively.
}

    \label{fig:1}
\end{figure}

To determine $N_{\alpha}$ self-consistently, one must appeal to the semiclassical kinetic framework, where the coupled dynamics of perturbed charge densities are governed by the continuity equation derived from the Boltzmann formalism \cite{PhysRevB.103.214310,Sukhachov2022PRL}. Using the current density in Eq.~\eqref{eq:Jalpha}, we express the continuity equation as
\begin{align}
    \label{eq:Kinetic}
   &\sum_\zeta\Bigg[T_{\alpha,\zeta}+D_\zeta\mathbf{\nabla}^2\delta_{\alpha,\zeta}-(i\vec{v_{\Omega,\zeta}}\cdot \vec{q}+i\omega)\delta_{\alpha,\zeta}\Bigg]N_\zeta=\notag\\
   &\hspace{3.2cm}-(e^2\nu_\alpha \vec{v}_{\Omega,\alpha}+i\sigma_\alpha(\omega) \vec{q})\cdot\vec{E}.
\end{align}
The first term on the left-hand side of Eq.~\eqref{eq:Kinetic}, $
T_{\alpha,\beta} = \delta_{\alpha,\beta}\sum_\zeta 1/\tau_{\alpha,\zeta} - 1/\tau_{\alpha,\beta}$,
describes internode scattering within the relaxation-time approximation. 
On the right-hand side, the first term represents the chiral anomaly source, while the remaining terms correspond to the conventional contributions from the continuity equation. For simplicity, we consider symmetric Weyl nodes, so that the transport properties are identical for both nodes. We can therefore suppress the node index, such that the intranode scattering time $\tau_{\alpha,\alpha}$ becomes $\tau$, while the internode scattering time $\tau_{\alpha,\beta}$ becomes $\tau_5$.
The analysis in the following focuses on the long-wavelength regime controlled by the critical momentum scale $q_c$, defined as 
$q_c \sim \frac{3}{v_{F,\alpha}^2 \tau \tau_5}$. This scale is derived comparing the diffusion term $q^2 D_\alpha$ with the internode scattering rate and the anomalous drift scale $\mathbf{v}_{\Omega,\alpha}\!\cdot\!\mathbf{q}$. We therefore restrict our attention to the regime $q \ll q_c$, where diffusion effects are negligible. In this limit, the dynamics of the valley-resolved densities $N_\alpha$ is governed predominantly by internode scattering and the anomaly-induced drift term associated with $\mathbf{v}_{\Omega,\alpha}$. In the collisionless limit, the kinetic equations for the node-resolved perturbed electron densities are decoupled. The inclusion of the internode scattering term couples the equations and introduces an imaginary contribution to the dielectric function, which accounts for the damping of the electromagnetic modes without modifying their dispersion at leading order. Under the above assumptions, we obtain the perturbed electron density imbalance in terms of the electric fields as
\begin{align}
 &-\frac{4\pi i \omega}{c^2}\sum_\zeta v_{\Omega,\zeta,x} N_\zeta=\lambda_{xx}E_x+\lambda_{xy}E_y,\label{lamx}\\
  &  -\frac{4\pi i \omega}{c^2}\sum_\zeta v_{\Omega,\zeta,y} N_\zeta=\lambda_{yx}E_x+\lambda_{yy}E_y,
    \label{lamy}
\end{align}
where the detailed derivation is provided in the Supplementary Material \cite{Supplemental}. The quantities $\lambda_{ij}$ in eq.\eqref{lamx} and eq.\eqref{lamy} describe the anomaly-induced electrodynamic response and are derived as follows

\begin{align}
\mathbf{[}\lambda\mathbf{]}_{\mathrm{ij}}
=\mathcal{A}\begin{pmatrix}
q^2_{TF}\,\sin^{2}(\theta)
&\frac{\sin(2\theta)}{2}\left( \tfrac{\omega_{p}^{2} q^{2}}{\omega^{2}} + q^2_{TF} \right)\
\\[3ex]
q^2_{TF}\,\frac{\sin(2\theta)}{2}\ 
&
\cos^{2}(\theta)\,\left( \tfrac{\omega_{p}^{2} q^{2}}{\omega^{2}} + q^2_{TF} \right)\ 
\end{pmatrix},
\label{eq: lamdaij}
\end{align}
 where $\mathcal{A}=\dfrac{ \omega^{2}v_{\Omega,\alpha}^{2} }
      { c^{2}\,\big(q^{2}v_{\Omega,\alpha}^{2}\cos^{2}(\theta) - \omega^{2}-i\frac{2\omega}{\tau_5}\big) }$ and $q_{TF}=\sqrt{4 \pi e^2\nu_0}$ is the Thomas-Fermi screening vector, which quantifies the electronic compressibility in the material. The inequality of the off-diagonal components $\lambda_{xy}$ and $\lambda_{yx}$ reflects the anisotropy introduced by propagation along the $y$ direction, which breaks the isotropy of the anomaly-induced response in both parallel and perpendicular configurations. Although intranode scattering and the associated diffusion are typically stronger in experiments,
internode scattering governs the relaxation and sets the finite lifetime of the collective
excitations in this parameter regime.
Diffusive effects become important only outside this region, at sufficiently large wave vectors. The crossover to diffusion-dominated dynamics is discussed separately in the next section and in the Supplemental Material \cite{Supplemental}. The expression in eq.\eqref{eq: lamdaij} represents 
the most general form of $\lambda_{ij}$ and incorporates the effects of internode scattering 
through the relaxation time $\tau_{5}$, which introduces an imaginary contribution to the 
dielectric response. In the collisionless regime, internode relaxation is negligible and 
Eq.~\eqref{eq: lamdaij} reduces to its real, non-dissipative form obtained by taking 
$1/\tau_{5}\to 0$. Retaining the internode scattering term restores the full expression, 
which we employ later to analyze damping and loss.
\paragraph*{Hybrid Bulk Plasmon Mode.—}
We next turn our attention to the analysis of the anomaly-induced electromagnetic response in the bulk 
and consider separately the cases of perpendicular and parallel configurations. We study the damping and loss peak of the bulk modes using the loss function 
$L(\omega,q) = -\Im[1/\varepsilon(\omega,q)]$, which directly probes the energy exchange between the electromagnetic field and the electronic degrees of freedom \cite{PhysRevB.91.035114}. The condition det $[\tilde{{\mathcal{M}}}^{\parallel/\perp}(\omega,q)] = 0$ yields the
full dispersion of the collective electromagnetic modes i.e.,
\begin{align}
    \underbrace{\bigg[\begin{pmatrix}
        \\&&\mathcal{M}^{\parallel/\perp}&&\vspace{3.5mm}
    \end{pmatrix}-\begin{pmatrix}
\lambda_{xx} & \lambda_{xy} &0\\[1ex]
\lambda_{yx} & \lambda_{yy}&0\\
0&0&0
\end{pmatrix}\bigg]}_{\tilde{{\mathcal{M}}}^{\parallel/\perp}}
    \begin{pmatrix}
        E_x\\
        E_y\\
        E_z
    \end{pmatrix}=
0.
    \label{General_maxwell_s}
\end{align}
In the perpendicular configuration, the electromagnetic modes decouple into two distinct sectors. The $E_z$ polarization sector corresponds to the conventional transverse bulk plasmon–polariton mode, while the $E_x$–$E_y$ sector describes coupled transverse polariton and longitudinal plasmon modes arising from the Weyl-node separation encoded in $\epsilon_2(\omega)$ and the chiral anomaly–induced terms $\lambda_{ij}$. Here, $\epsilon_2(\omega)=\epsilon_\infty\omega_b/\omega$, with $\omega_b = \frac{2e^2|\bm b|}{\pi\hbar\epsilon_\infty}$. Accordingly, in what follows we focus on the coupled $x$--$y$ sector by evaluating the 
dielectric response obtained from the corresponding $2\times2$ sub-determinant of 
Eq.\eqref{General_maxwell_s}, which yields three modes in total: two gapped modes and a linearly dispersing 
anomaly-induced branch $\omega_{\mathrm{anom}} = uq$ in the long-wavelength limit 
$(q\!\to\!0)$ as shown in fig. \ref{fig:1a}. The velocity $u$ is given by
\begin{equation}
    u=\sqrt{
\frac{
v_{\Omega,\alpha}^{2}\omega_p^{2}(\epsilon_\infty-1)\cos^{2}\theta}{
\epsilon_\infty\omega_p^{2}+v_{\Omega,\alpha}^{2}q_{TF}^2
}},
\end{equation}
where $\epsilon_\infty$ is the background dielectric constant of the material. At finite momentum, this branch hybridizes with one of the gapped modes, 
producing an avoided crossing and a hybrid bulk excitation arising from the 
anomaly-induced mixing of charge oscillations and oscillations in the chiral charge imbalance between Weyl valleys of opposite chirality. This behavior persists for all propagation angles $0 \le \theta < \pi/2$, as shown in fig. \ref{fig:1a}. In contrast, for $\theta=\pi/2$, corresponding to the magnetic field perpendicular to the propagation, the anomaly-induced coupling vanishes and neither the linear mode nor the hybridization is present. In the long-wavelength limit ($q \to 0$), the velocity of the linear mode, $u$, is identical for both parallel and perpendicular configurations. This is because the contribution from $\epsilon_2(\omega)$, which encodes the Weyl-node separation, is negligible in leading order since it scales as $\sim q^2$, making the distinction between configurations irrelevant for the linear-mode velocity. For a magnetic field $\mathbf{B}_0$ lying in the $y$–$z$ plane, the bulk response remains coupled, giving rise to four modes: three gapped modes and one linearly dispersing gapless mode as shown in fig. \ref{fig:1b}. Nevertheless, the long-wavelength linear mode and its hybridization with one of the low-energy gapped modes remain unaffected. Although the dispersion relation admits this mode, it does not appear as a distinct peak in the loss spectrum, indicating that it is strongly damped and does not form a well-defined excitation as shown in fig. \ref{fig:1b}. We now turn to the parallel configuration. For arbitrary $\theta \in [0,\pi/2]$, all bulk electromagnetic modes are coupled, giving rise to three gapped modes~\cite{pellegrino2015helicons} and a gapless linear mode, $\omega_{\mathrm{anom}} = uq$, in the long-wavelength limit $(q \!\to\! 0)$. While this mode hybridizes with the lowest-energy gapped bulk mode at finite $q$, it does not form a distinct hybrid excitation in the loss spectrum, as shown in Fig.~\ref{fig:1c}.
For $\theta=0$, the bulk modes decouple into a longitudinal plasmon sector and a transverse plasmon-polariton sector (discussed in the Supplemental Material~\cite{Supplemental}). Since the transverse sector remains unaffected by the anomaly, it is not shown in Fig.~\ref{fig:1d}. The figure displays the anomaly-modified longitudinal sector, where the anomaly-induced coupling hybridizes a linear gapless mode with the longitudinal bulk plasmon mode, producing a well-defined and weakly damped hybrid plasmon mode visible as a sharp blue ridge in the loss spectrum. In contrast, for $\theta=\pi/2$, the linear gapless mode and the hybridization are absent, similar to the perpendicular configuration. In the parallel configuration, the system is isotropic in the plane along the direction of the wave-vector (see supplementary material \cite{Supplemental} for detailed discussion). Consequently, the physics is unchanged whether $\mathbf{B}_0$ lies in the $x$-$y$ or $y$-$z$ plane, making further distinction unnecessary. 

To obtain quantitative estimates, we use representative parameters of a Weyl semimetal. 
Experimental realizations typically exhibit Fermi energies measured from the Weyl nodes 
in the range $\mu \sim 10\text{--}60\,\mathrm{meV}$ \cite{Huang2015,Arnold2016,PhysRevB.97.235416,Liu2019}, with reported Fermi velocities 
$v_F \sim (3\text{--}9)\times10^5\,\mathrm{m/s}$ (see Refs.~\cite{PhysRevB.92.241108,PhysRevLett.114.117201,Liang2015}). 
For these parameters, the semiclassical description remains valid provided 
$\mu \gtrsim \hbar \omega_c$. Depending on the specific values of $\mu$ and $v_F$, 
this requirement is satisfied for magnetic fields up to a few Tesla \cite{Zhang2016,Potter2014}. Accordingly, 
in our calculations we consider magnetic fields in the approximate range 
$B \sim 0.1\text{--}4\,\mathrm{T}$, where the semiclassical treatment remains applicable 
for the relevant doping levels. All numerical results presented below are obtained within this experimentally relevant 
parameter regime. With these values, the anomaly-induced velocity reaches 
$v_{\Omega,\alpha} \sim 0.05\,c$, which is sufficient to generate the anomaly-induced 
hybrid collective modes discussed in this work. Internode scattering is incorporated 
through a finite intervalley relaxation rate $1/\tau_5$. In the calculations we take 
$\omega_b = 0.5\,\omega_p$ \cite{PhysRevB.93.241402} and $1/\tau_5 = 0.001\,\omega_p$ \cite{PhysRevB.103.214310}, consistent with 
experimentally inferred intervalley relaxation times in Weyl semimetals and 
corresponding to weak damping of the collective modes. We set $\epsilon_\infty = 13$ \cite{PhysRevB.92.241108}. The low Fermi energies in Weyl semimetals ($\mu \sim 10$–$60~\mathrm{meV}$) unlike conventional metals yield a small Thomas–Fermi wavevector, which scales with the density of states at the Fermi level, leading to weak screening and allowing interband effects to be neglected in the regime of interest.

We now clarify the regime in which diffusive dynamics may be neglected in favor of
internode relaxation.
Diffusion enters the density response through terms proportional to $D q^{2}$, whereas
internode scattering is governed by the eigenvalues of the collision matrix
$T_{\alpha\beta}$.
Their competition defines a characteristic wave vector
$q_{c}$ as discussed above. The anomaly-induced hybridization occurs at a characteristic wave vector
$q_{h} = \omega_{p}/u$, determined by the crossing of the linear electromagnetic mode
$\omega = u q$ with the longitudinal plasmon branch.
Requiring the hybridization to take place within the collisionless regime,
$q_{h} < q_{c}$, leads to the condition 
\begin{equation}
    \eta<1,
    \label{eq: hyb_cond}
\end{equation}
where $\eta=\bigg(\frac{v^2_{\Omega,\alpha}q^2_{TF}+\epsilon_\infty\omega_p^2}{v^2_{\Omega,\alpha,y}(\epsilon_\infty-1)}\bigg)\Bigg/(3/v^2_{F,\alpha}\tau_5\tau).$ In inversion symmetry broken Weyl semimetals such as TaAs and related compounds, Weyl nodes are typically separated by large momenta in the Brillouin zone \cite{PhysRevX.5.031013,Huang2015,PhysRevX.5.011029}. As a result, intervalley scattering requires large momentum transfer and is strongly suppressed in the presence of smooth disorder, leading to $\tau_5 \gg \tau$. In contrast, in time reversal symmetry broken Weyl semimetals or in field-induced Weyl phases from Dirac semimetals \cite{PhysRevB.97.235416,PhysRevB.96.155141,Zhong2025,PhysRevB.98.075123,PhysRevB.88.165105}, the node separation can be smaller, allowing comparatively stronger internode scattering. Motivated by these considerations, we adopt $\tau_5/\tau \sim \mathcal{O}(10)$ as a representative value. This choice satisfies the condition in Eq.~\eqref{eq: hyb_cond} over the relevant parameter regime (see Supplemental Material \cite{Supplemental} for a detailed discussion).

\paragraph*{Summary and Conclusion.—}
We report a \emph{previously unidentified bulk collective mode} in doped Weyl semimetals: an \emph{Anomaly-induced hybrid electromagnetic mode} generated by the interplay of chiral anomaly and its associated chiral magnetic current along the direction of the wave-vector. This excitation is qualitatively distinct from previously known anomaly-related collective excitations. Unlike chiral magnetic wave (CMW)~\cite{Kharzeev2010CMW,PhysRevB.92.115310}, and chiral zero sound (CZS)~\cite{PhysRevX.9.021053}, the present excitation is an intrinsically electromagnetic hybrid mode in the presence of a weak magnetic field in the direction of wave-vector. Its origin lies in the coupling between plasmonic response of the Weyl medium  and the anomaly-induced oscillations in chemical potential imbalance when the magnetic field is along the propagation, and it is distinct from previously discussed chiral magnetic plasmons~\cite{Gorbar2017CMP}, which remain gapped plasma oscillations. Our analysis further reveals a pronounced dependence of the hybrid mode on the relative orientation between the direction of the wave-vector and the chiral magnetic response.
In the perpendicular configuration, an anomaly-induced hybrid branch exists for all
angles $0 \le \theta < \pi/2$, exhibiting a robust low-momentum linear dispersion and
hybridizing with the longitudinal plasmon at finite wave vectors through an avoided
level crossing.
The linear hybrid mode is absent despite the presence of chiral anomaly when the associated chiral magnetic response is oriented perpendicular $(\theta=\pi/2)$ to the direction of the wave-vector.
In the parallel configuration, the behavior is more restrictive: for $\theta=0$ the
anomaly modifies the longitudinal plasmon sector, yielding a well-defined and weakly
damped collective excitation. However, for any oblique angle $0<\theta<\pi/2$ the resulting hybrid mode becomes strongly
damped due to internode relaxation.
More broadly, the present work demonstrates that treating the chiral anomaly dynamically, together with the orientation of its associated chiral magnetic current relative to the direction of the wave-vector, fundamentally alters the electromagnetic response of Weyl semimetals. This dynamical treatment leads to a
hybridization between charge density oscillations and fluctuations of the chemical
potential imbalance between Weyl nodes of opposite chirality, giving rise to an
anomaly induced hybrid collective mode. Our work establishes a framework for exploring a wide class of anomaly-driven optical and electromagnetic phenomena in Weyl semimetals by treating the chiral anomaly dynamically and accounting for the orientation of its associated chiral magnetic response. Recent studies have revealed further consequences of band tilting in the electromagnetic response of Weyl semimetals, including tilt-induced surface plasmon-polariton modes~\cite{Jafaria} and vortical responses generated by inhomogeneous tilt textures. The curl of the tilt velocity can also act as an effective magnetic field and generate a chiral vortical velocity analogous to the anomaly-induced chiral magnetic velocity considered here~\cite{Jafarib}, it would be interesting, as future work, to explore whether similar hybrid collective electromagnetic modes emerge in such systems.

\paragraph*{Data Availability Statement.—}
The data that support the findings of this study are available from the corresponding author upon reasonable request~\cite{DataAvailability}.
\bibliographystyle{apsrev4-2}
\bibliography{references}

@Article{Weyl1929,
author={Weyl, Hermann},
title={Elektron und Gravitation. I},
journal={Zeitschrift f{\"u}r Physik},
year={1929},
month={May},
day={01},
volume={56},
number={5},
pages={330-352},
issn={0044-3328},
doi={10.1007/BF01339504},
url={https://doi.org/10.1007/BF01339504}
}

@article{Balents2011,
  author  = {L. Balents},
  title   = {Weyl electrons kiss},
  journal = {Physics},
  volume  = {4},
  pages   = {36},
  year    = {2011},
  doi     = {10.1103/Physics.4.36}
}

@article{Wan2011,
  title={Topological semimetal and Fermi-arc surface states in the electronic structure of pyrochlore iridates},
  author={Wan, X. and Turner, A. M. and Vishwanath, A. and Savrasov, S. Y.},
  journal={Physical Review B},
  volume={83},
  number={20},
  pages={205101},
  year={2011}
}

@article{Burkov2011,
  title={Weyl semimetal in a topological insulator multilayer},
  author={Burkov, A. A. and Balents, L.},
  journal={Physical Review Letters},
  volume={107},
  number={12},
  pages={127205},
  year={2011}
}

@article{Armitage2018,
  title={Weyl and Dirac semimetals in three-dimensional solids},
  author={Armitage, N. P. and Mele, E. J. and Vishwanath, A.},
  journal={Reviews of Modern Physics},
  volume={90},
  number={1},
  pages={015001},
  year={2018}
}

@article{Murakami2007,
  author  = {S. Murakami},
  title   = {Phase transition between the quantum spin Hall and insulator phases in 3D: Emergence of a topological gapless phase},
  journal = {New J. Phys.},
  volume  = {9},
  pages   = {356},
  year    = {2007},
  doi     = {10.1088/1367-2630/9/9/356}
}

@article{Burkov2014,
  author  = {A. A. Burkov},
  title   = {Anomalous Hall Effect in Weyl Metals},
  journal = {Phys. Rev. Lett.},
  volume  = {113},
  pages   = {187202},
  year    = {2014},
  doi     = {10.1103/PhysRevLett.113.187202}
}

@article{Shekhar2018,
  author  = {Shekhar, Chandra and Kumar, Nitesh and Grinenko, V. and Singh, Sanjay and Sarkar, R. and Luetkens, H. and Wu, Shu-Chun and Zhang, Yang and Komarek, Alexander C. and Kampert, Erik and Skourski, Yurii and Wosnitza, Jochen and Schnelle, Walter and McCollam, Alix and Zeitler, Uli and Kübler, Jürgen and Yan, Binghai and Klauss, H.-H. and Parkin, S. S. P. and Felser, Claudia},
  title   = {Anomalous Hall effect in Weyl semimetal half-Heusler compounds RPtBi (R = Gd and Nd)},
  journal = {Proc. Natl. Acad. Sci. U.S.A.},
  volume  = {115},
  number  = {37},
  pages   = {9140--9144},
  year    = {2018},
  doi     = {10.1073/pnas.1810842115}
}

@article{Bednik2016,
  author  = {Bednik, G. and Zyuzin, A. A. and Burkov, A. A.},
  title   = {Anomalous Hall effect in Weyl superconductors},
  journal = {New J. Phys.},
  volume  = {18},
  pages   = {085002},
  year    = {2016},
  doi     = {10.1088/1367-2630/18/8/085002}
}

@article{Kargarian2015,
  author = {M. Kargarian and M. Randeria and N. Trivedi},
  title = {Theory of Kerr and Faraday rotations and linear dichroism in topological Weyl semimetals},
  journal = {Sci. Rep.},
  volume = {5},
  pages = {12683},
  year = {2015},
  doi = {10.1038/srep12683}
}

@article{Ghosh2023,
  author = {S. Ghosh and A. Sahoo and S. Nandy},
  title = {Theoretical investigations on Kerr and Faraday rotations in topological multi-Weyl semimetals},
  journal = {SciPost Phys.},
  volume = {15},
  pages = {133},
  year = {2023},
  doi = {10.21468/SciPostPhys.15.4.133}
}

@article{Adler1969,
  author = {S. L. Adler},
  title = {Axial-vector vertex in spinor electrodynamics},
  journal = {Phys. Rev.},
  volume = {177},
  pages = {2426},
  year = {1969},
  doi = {10.1103/PhysRev.177.2426}
}

@article{Bell1969,
  author = {J. S. Bell and R. Jackiw},
  title = {A PCAC puzzle: $\pi^0 \rightarrow \gamma \gamma$ in the $\sigma$ model},
  journal = {Nuovo Cimento A},
  volume = {60},
  pages = {47},
  year = {1969},
  doi = {10.1007/BF02823296}
}

@article{Son2013,
  title={Chiral anomaly and classical negative magnetoresistance of Weyl metals},
  author={Son, D. T. and Spivak, B. Z.},
  journal={Physical Review B},
  volume={88},
  number={10},
  pages={104412},
  year={2013}
}

@article{dosReis2016,
  author = {R. D. dos Reis and M. O. Ajeesh and N. Kumar and F. Arnold and C. Shekhar and M. Naumann and M. Schmidt and M. Nicklas and E. Hassinger},
  title = {On the search for the chiral anomaly in Weyl semimetals: the negative longitudinal magnetoresistance},
  journal = {New J. Phys.},
  volume = {18},
  pages = {085006},
  year = {2016},
  doi = {10.1088/1367-2630/18/8/085006}
}

@article{PhysRevB.99.075114,
  title = {Chiral anomaly in type-I Weyl semimetals: Comprehensive analysis within a semiclassical Fermi surface harmonics approach},
  author = {Johansson, Annika and Henk, J\"urgen and Mertig, Ingrid},
  journal = {Phys. Rev. B},
  volume = {99},
  issue = {7},
  pages = {075114},
  numpages = {12},
  year = {2019},
  month = {Feb},
  publisher = {American Physical Society},
  doi = {10.1103/PhysRevB.99.075114},
  url = {https://link.aps.org/doi/10.1103/PhysRevB.99.075114}
}

@article{Fukushima2008,
  author = {K. Fukushima and D. E. Kharzeev and H. J. Warringa},
  title = {Chiral magnetic effect},
  journal = {Phys. Rev. D},
  volume = {78},
  pages = {074033},
  year = {2008},
  doi = {10.1103/PhysRevD.78.074033}
}

@article{Kharzeev2014,
  author = {D. E. Kharzeev},
  title = {The chiral magnetic effect and anomaly-induced transport},
  journal = {Prog. Part. Nucl. Phys.},
  volume = {75},
  pages = {133},
  year = {2014},
  doi = {10.1016/j.ppnp.2014.01.002}
}

@article{Li2016,
  author = {Q. Li and D. E. Kharzeev and C. Zhang and Y. Huang and I. Pletikosic and A. V. Fedorov and R. D. Zhong and J. A. Schneeloch and G. D. Gu and T. Valla},
  title = {Chiral magnetic effect in ZrTe5},
  journal = {Nat. Phys.},
  volume = {12},
  pages = {550},
  year = {2016},
  doi = {10.1038/nphys3648}
}

@article{PhysRevB.93.201202,
  title = {Photovoltaic chiral magnetic effect in Weyl semimetals},
  author = {Taguchi, Katsuhisa and Imaeda, Tatsushi and Sato, Masatoshi and Tanaka, Yukio},
  journal = {Phys. Rev. B},
  volume = {93},
  issue = {20},
  pages = {201202},
  numpages = {5},
  year = {2016},
  month = {May},
  publisher = {American Physical Society},
  doi = {10.1103/PhysRevB.93.201202},
  url = {https://link.aps.org/doi/10.1103/PhysRevB.93.201202}
}

@article{Burkov2015,
  title = {Negative longitudinal magnetoresistance in Dirac and Weyl metals},
  author = {Burkov, A. A.},
  journal = {Phys. Rev. B},
  volume = {91},
  pages = {245157},
  year = {2015},
  doi = {10.1103/PhysRevB.91.245157}
}

@article{Nandy2017,
  title={Chiral anomaly as the origin of the planar Hall effect in Weyl semimetals},
  author={Nandy, S. and Sharma, G. and Taraphder, A. and Tewari, S.},
  journal={Physical Review Letters},
  volume={119},
  number={17},
  pages={176804},
  year={2017}
}

@article{Zhang2021,
  author = {C.-L. Zhang and T. Liang and M. S. Bahramy and N. Ogawa and V. Kocsis and K. Ueda and Y. Kaneko and M. Kriener and Y. Tokura},
  title = {Berry curvature generation detected by Nernst responses in ferroelectric Weyl semimetal},
  journal = {Proc. Natl. Acad. Sci. U.S.A.},
  volume = {118},
  pages = {e2111855118},
  year = {2021},
  doi = {10.1073/pnas.2111855118}
}

@article{Lv2013,
  author = {M. Lv and S.-C. Zhang},
  title = {Dielectric function, Friedel oscillation and plasmons in Weyl semimetals},
  journal = {Int. J. Mod. Phys. B},
  volume = {27},
  pages = {1350177},
  year = {2013},
  doi = {10.1142/S0217979213501774}
}

@article{pellegrino2015helicons,
  title={Helicons in Weyl semimetals},
  author={Pellegrino, F. M. D. and Principi, A. and Vignale, G. and Polini, M.},
  journal={Physical Review B},
  volume={92},
  number={12},
  pages={125411},
  year={2015}
}

@article{PhysRevB.104.205141,
  title = {Three-dimensional topological plasmons in Weyl semimetals},
  author = {Zhang, Furu and Gao, Yang and Zhang, Wei},
  journal = {Phys. Rev. B},
  volume = {104},
  issue = {20},
  pages = {205141},
  numpages = {12},
  year = {2021},
  month = {Nov},
  publisher = {American Physical Society},
  doi = {10.1103/PhysRevB.104.205141},
  url = {https://link.aps.org/doi/10.1103/PhysRevB.104.205141}
}

@article{PhysRevB.99.075137,
  title = {Optical properties and electromagnetic modes of Weyl semimetals},
  author = {Chen, Qianfan and Kutayiah, A. Ryan and Oladyshkin, Ivan and Tokman, Mikhail and Belyanin, Alexey},
  journal = {Phys. Rev. B},
  volume = {99},
  issue = {7},
  pages = {075137},
  numpages = {24},
  year = {2019},
  month = {Feb},
  publisher = {American Physical Society},
  doi = {10.1103/PhysRevB.99.075137},
  url = {https://link.aps.org/doi/10.1103/PhysRevB.99.075137}
}

@article{PhysRevLett.120.037403,
  title = {Magnetopolaritons in Weyl Semimetals in a Strong Magnetic Field},
  author = {Long, Zhongqu and Wang, Yongrui and Erukhimova, Maria and Tokman, Mikhail and Belyanin, Alexey},
  journal = {Phys. Rev. Lett.},
  volume = {120},
  issue = {3},
  pages = {037403},
  numpages = {5},
  year = {2018},
  month = {Jan},
  publisher = {American Physical Society},
  doi = {10.1103/PhysRevLett.120.037403},
  url = {https://link.aps.org/doi/10.1103/PhysRevLett.120.037403}
}

@article{PhysRevB.91.035114,
  title = {Plasmon mode as a detection of the chiral anomaly in Weyl semimetals},
  author = {Zhou, Jianhui and Chang, Hao-Ran and Xiao, Di},
  journal = {Phys. Rev. B},
  volume = {91},
  issue = {3},
  pages = {035114},
  numpages = {8},
  year = {2015},
  month = {Jan},
  publisher = {American Physical Society},
  doi = {10.1103/PhysRevB.91.035114},
  url = {https://link.aps.org/doi/10.1103/PhysRevB.91.035114}
}

@article{PhysRevLett.118.127601,
  title = {Consistent Chiral Kinetic Theory in Weyl Materials: Chiral Magnetic Plasmons},
  author = {Gorbar, E. V. and Miransky, V. A. and Shovkovy, I. A. and Sukhachov, P. O.},
  journal = {Phys. Rev. Lett.},
  volume = {118},
  issue = {12},
  pages = {127601},
  numpages = {6},
  year = {2017},
  month = {Mar},
  publisher = {American Physical Society},
  doi = {10.1103/PhysRevLett.118.127601},
  url = {https://link.aps.org/doi/10.1103/PhysRevLett.118.127601}
}

@article{agarwal2020novel,
  title={Novel plasmon mode in tilted Dirac semimetals},
  author={Agarwal, A. and Mishra, S.},
  journal={Physical Review B},
  volume={101},
  number={4},
  pages={045417},
  year={2020}
}

@article{PhysRevB.103.214310,
  title = {Anomalous sound attenuation in Weyl semimetals in magnetic and pseudomagnetic fields},
  author = {Sukhachov, P. O. and Glazman, L. I.},
  journal = {Phys. Rev. B},
  volume = {103},
  issue = {21},
  pages = {214310},
  numpages = {17},
  year = {2021},
  month = {Jun},
  publisher = {American Physical Society},
  doi = {10.1103/PhysRevB.103.214310},
  url = {https://link.aps.org/doi/10.1103/PhysRevB.103.214310}
}

@article{Sukhachov2022PRL,
  title = {Anomalous Electromagnetic Field Penetration in a Weyl or Dirac Semimetal},
  author = {Sukhachov, P. O. and Glazman, L. I.},
  journal = {Phys. Rev. Lett.},
  volume = {128},
  issue = {14},
  pages = {146801},
  numpages = {7},
  year = {2022},
  month = {Apr},
  publisher = {American Physical Society},
  doi = {10.1103/PhysRevLett.128.146801},
  url = {https://link.aps.org/doi/10.1103/PhysRevLett.128.146801}
}

@article{Huang2015,
  author = {X. Huang and L. Zhao and Y. Long and P. Wang and D. Chen and Z. Yang and H. Liang and M. Xue and H. Weng and Z. Fang and X. Dai and G. Chen},
  title = {Observation of the Chiral-Anomaly-Induced Negative Magnetoresistance in 3D Weyl Semimetal TaAs},
  journal = {Phys. Rev. X},
  volume = {5},
  pages = {031023},
  year = {2015},
  doi = {10.1103/PhysRevX.5.031023}
}

@article{Arnold2016,
  author = {F. Arnold and M. Naumann and S.-C. Wu and Y. Sun and M. Schmidt and H. Borrmann and C. Felser and B. Yan and E. Hassinger},
  title = {Chiral Weyl Pockets and Fermi Surface Topology of the Weyl Semimetal TaAs},
  journal = {Phys. Rev. Lett.},
  volume = {117},
  pages = {146401},
  year = {2016},
  doi = {10.1103/PhysRevLett.117.146401}
}

@article{PhysRevB.97.235416,
  title = {Topological surface Fermi arcs in the magnetic Weyl semimetal ${\mathrm{Co}}_{3}{\mathrm{Sn}}_{2}{\mathrm{S}}_{2}$},
  author = {Xu, Qiunan and Liu, Enke and Shi, Wujun and Muechler, Lukas and Gayles, Jacob and Felser, Claudia and Sun, Yan},
  journal = {Phys. Rev. B},
  volume = {97},
  issue = {23},
  pages = {235416},
  numpages = {8},
  year = {2018},
  month = {Jun},
  publisher = {American Physical Society},
  doi = {10.1103/PhysRevB.97.235416},
  url = {https://link.aps.org/doi/10.1103/PhysRevB.97.235416}
}

@article{Liu2019,
  author = {D. F. Liu and A. J. Liang and E. K. Liu and Q. N. Xu and Y. W. Li and C. Chen and D. Pei and W. J. Shi and S. K. Mo and Y. L. Chen and Z. K. Liu},
  title = {Magnetic Weyl semimetal phase in a Kagome crystal},
  journal = {Science},
  volume = {365},
  pages = {1282},
  year = {2019},
  doi = {10.1126/science.aav2873}
}

@article{PhysRevLett.114.117201,
  title = {Linear Magnetoresistance Caused by Mobility Fluctuations in $n$-Doped ${\mathrm{Cd}}_{3}{\mathrm{As}}_{2}$},
  author = {Narayanan, A. and Watson, M. D. and Blake, S. F. and Bruyant, N. and Drigo, L. and Chen, Y. L. and Prabhakaran, D. and Yan, B. and Felser, C. and Kong, T. and Canfield, P. C. and Coldea, A. I.},
  journal = {Phys. Rev. Lett.},
  volume = {114},
  issue = {11},
  pages = {117201},
  numpages = {5},
  year = {2015},
  month = {Mar},
  publisher = {American Physical Society},
  doi = {10.1103/PhysRevLett.114.117201},
  url = {https://link.aps.org/doi/10.1103/PhysRevLett.114.117201}
}

@Article{Liang2015,
author={Liang, Tian
and Gibson, Quinn
and Ali, Mazhar N.
and Liu, Minhao
and Cava, R. J.
and Ong, N. P.},
title={Ultrahigh mobility and giant magnetoresistance in the Dirac semimetal Cd3As2},
journal={Nature Materials},
year={2015},
month={Mar},
day={01},
volume={14},
number={3},
pages={280-284},
issn={1476-4660},
doi={10.1038/nmat4143},
url={https://doi.org/10.1038/nmat4143}
}

@article{PhysRevB.93.241402,
  title = {Surface plasmon polaritons in topological Weyl semimetals},
  author = {Hofmann, Johannes and Das Sarma, Sankar},
  journal = {Phys. Rev. B},
  volume = {93},
  issue = {24},
  pages = {241402},
  numpages = {5},
  year = {2016},
  month = {Jun},
  publisher = {American Physical Society},
  doi = {10.1103/PhysRevB.93.241402},
  url = {https://link.aps.org/doi/10.1103/PhysRevB.93.241402}
}

@article{PhysRevB.92.241108,
  title = {Optical evidence for a Weyl semimetal state in pyrochlore ${\mathrm{Eu}}_{2}{\mathrm{Ir}}_{2}{\mathrm{O}}_{7}$},
  author = {Sushkov, A. B. and Hofmann, J. B. and Jenkins, G. S. and Ishikawa, J. and Nakatsuji, S. and Das Sarma, S. and Drew, H. D.},
  journal = {Phys. Rev. B},
  volume = {92},
  issue = {24},
  pages = {241108},
  numpages = {4},
  year = {2015},
  month = {Dec},
  publisher = {American Physical Society},
  doi = {10.1103/PhysRevB.92.241108},
  url = {https://link.aps.org/doi/10.1103/PhysRevB.92.241108}
}

@article{Potter2014,
  author = {A. C. Potter and I. Kimchi and A. Vishwanath},
  title = {Quantum oscillations from surface Fermi arcs in Weyl and Dirac semimetals},
  journal = {Nat. Commun.},
  volume = {5},
  pages = {5161},
  year = {2014},
  doi = {10.1038/ncomms6161}
}

@article{Zhang2016,
  author = {Y. Zhang and D. Bulmash and P. Hosur and A. C. Potter and A. Vishwanath},
  title = {Quantum oscillations from generic surface Fermi arcs and bulk chiral modes in Weyl semimetals},
  journal = {Sci. Rep.},
  volume = {6},
  pages = {23741},
  year = {2016},
  doi = {10.1038/srep23741}
}

@article{PhysRevX.5.031013,
  title = {Experimental Discovery of Weyl Semimetal TaAs},
  author = {Lv, B. Q. and Weng, H. M. and Fu, B. B. and Wang, X. P. and Miao, H. and Ma, J. and Richard, P. and Huang, X. C. and Zhao, L. X. and Chen, G. F. and Fang, Z. and Dai, X. and Qian, T. and Ding, H.},
  journal = {Phys. Rev. X},
  volume = {5},
  issue = {3},
  pages = {031013},
  numpages = {8},
  year = {2015},
  month = {Jul},
  publisher = {American Physical Society},
  doi = {10.1103/PhysRevX.5.031013},
  url = {https://link.aps.org/doi/10.1103/PhysRevX.5.031013}
}

@article{PhysRevX.5.011029,
  title = {Weyl Semimetal Phase in Noncentrosymmetric Transition-Metal Monophosphides},
  author = {Weng, Hongming and Fang, Chen and Fang, Zhong and Bernevig, B. Andrei and Dai, Xi},
  journal = {Phys. Rev. X},
  volume = {5},
  issue = {1},
  pages = {011029},
  numpages = {10},
  year = {2015},
  month = {Mar},
  publisher = {American Physical Society},
  doi = {10.1103/PhysRevX.5.011029},
  url = {https://link.aps.org/doi/10.1103/PhysRevX.5.011029}
}

@article{PhysRevB.96.155141,
  title = {Weyl semimetal induced from a Dirac semimetal by magnetic doping},
  author = {Deng, Ming-Xun and Luo, Wei and Wang, Rui-Qiang and Sheng, L. and Xing, D. Y.},
  journal = {Phys. Rev. B},
  volume = {96},
  issue = {15},
  pages = {155141},
  numpages = {7},
  year = {2017},
  month = {Oct},
  publisher = {American Physical Society},
  doi = {10.1103/PhysRevB.96.155141},
  url = {https://link.aps.org/doi/10.1103/PhysRevB.96.155141}
}

@Article{Zhong2025,
author={Zhong, Jingyuan
and Wang, Jianfeng
and Yang, Ming
and Liu, Jie
and Ren, Zhizhen
and Huang, Anping
and Shi, Zhixiang
and Zhu, Zengwei
and Shi, Yan
and Hao, Weichang
and Zhuang, Jincheng
and Du, Yi},
title={Field manipulation of Weyl modes in an ideal Dirac semimetal},
journal={Nature Communications},
year={2025},
month={Nov},
day={28},
volume={16},
number={1},
pages={10785},
issn={2041-1723},
doi={10.1038/s41467-025-65832-7},
url={https://doi.org/10.1038/s41467-025-65832-7}
}

@article{PhysRevB.98.075123,
  title = {Magnetic field induced Weyl semimetal from Wannier-function-based tight-binding model},
  author = {Villanova, John W. and Park, Kyungwha},
  journal = {Phys. Rev. B},
  volume = {98},
  issue = {7},
  pages = {075123},
  numpages = {14},
  year = {2018},
  month = {Aug},
  publisher = {American Physical Society},
  doi = {10.1103/PhysRevB.98.075123},
  url = {https://link.aps.org/doi/10.1103/PhysRevB.98.075123}
}

@article{PhysRevB.88.165105,
  title = {Engineering Weyl nodes in Dirac semimetals by a magnetic field},
  author = {Gorbar, E. V. and Miransky, V. A. and Shovkovy, I. A.},
  journal = {Phys. Rev. B},
  volume = {88},
  issue = {16},
  pages = {165105},
  numpages = {6},
  year = {2013},
  month = {Oct},
  publisher = {American Physical Society},
  doi = {10.1103/PhysRevB.88.165105},
  url = {https://link.aps.org/doi/10.1103/PhysRevB.88.165105}
}

@article{Kharzeev2010CMW,
  title={Chiral magnetic wave},
  author={Kharzeev, Dmitri E. and Yee, Ho-Ung},
  journal={Phys. Rev. D},
  volume={83},
  pages={085007},
  year={2011},
  doi={10.1103/PhysRevD.83.085007}
}

@article{PhysRevB.92.115310,
  title = {Chiral electromagnetic waves in Weyl semimetals},
  author = {Zyuzin, Alexander A. and Zyuzin, Vladimir A.},
  journal = {Phys. Rev. B},
  volume = {92},
  issue = {11},
  pages = {115310},
  numpages = {4},
  year = {2015},
  month = {Sep},
  publisher = {American Physical Society},
  doi = {10.1103/PhysRevB.92.115310},
  url = {https://link.aps.org/doi/10.1103/PhysRevB.92.115310}
}

@article{PhysRevX.9.021053,
  title = {Hear the Sound of Weyl Fermions},
  author = {Song, Zhida and Dai, Xi},
  journal = {Phys. Rev. X},
  volume = {9},
  issue = {2},
  pages = {021053},
  numpages = {22},
  year = {2019},
  month = {Jun},
  publisher = {American Physical Society},
  doi = {10.1103/PhysRevX.9.021053},
  url = {https://link.aps.org/doi/10.1103/PhysRevX.9.021053}
}

@article{Gorbar2017CMP,
  title={Chiral magnetic plasmons in anomalous relativistic matter},
  author={Gorbar, E. V. and Miransky, V. A. and Shovkovy, I. A. and Sukhachov, P. O.},
  journal={Phys. Rev. B},
  volume={95},
  pages={115202},
  year={2017},
  doi={10.1103/PhysRevB.95.115202}
}

@misc{Supplemental,
    note = {See Supplemental Material for the derivations of the Maxwell wave equation and kinetic equation in the presence of weak magnetic field, derivation of anomaly-induced terms $\lambda$, detailed discussion of the validity.}
}

@article{Jafaric,
  title = {Sound of Fermi arcs: a linearly dispersing gapless surface plasmon mode in undoped Weyl semimetals},
  author = {Adinehvand, F. and Faraei, Z. and Farajollahpour, T. and Jafari, S. A.},
  journal = {Phys. Rev. B},
  volume = {100},
  issue = {19},
  pages = {195408},
  numpages = {10},
  year = {2019},
  month = {Nov},
  publisher = {American Physical Society},
  doi = {10.1103/PhysRevB.100.195408},
  url = {https://link.aps.org/doi/10.1103/PhysRevB.100.195408}
}

@article{Jafaria,
  title = {Electrodynamics of tilted Dirac and Weyl materials: A unique platform for unusual surface plasmon polaritons},
  author = {Jalali-Mola, Z. and Jafari, S. A.},
  journal = {Phys. Rev. B},
  volume = {100},
  issue = {20},
  pages = {205413},
  numpages = {15},
  year = {2019},
  month = {Nov},
  publisher = {American Physical Society},
  doi = {10.1103/PhysRevB.100.205413},
  url = {https://link.aps.org/doi/10.1103/PhysRevB.100.205413}
}

@article{Jafarib,
  title = {Tilt-induced vortical response and mixed anomaly in inhomogeneous Weyl matter},
  author = {Rostamzadeh, Saber and Tasdemir, Sevval and Sarisaman, Mustafa and Jafari, S. A. and Goerbig, Mark-Oliver},
  journal = {Phys. Rev. B},
  volume = {107},
  issue = {7},
  pages = {075155},
  numpages = {10},
  year = {2023},
  month = {Feb},
  publisher = {American Physical Society},
  doi = {10.1103/PhysRevB.107.075155},
  url = {https://link.aps.org/doi/10.1103/PhysRevB.107.075155}
}

@misc{DataAvailability,
  note = {Data supporting the findings of this study are available from the corresponding author upon reasonable request}
}

\end{document}